\documentclass{article}

\usepackage{arxiv}

\usepackage[utf8]{inputenc} 
\usepackage[T1]{fontenc}    
\usepackage{hyperref}       
\usepackage{url}            
\usepackage{booktabs}       
\usepackage{amsfonts}       
\usepackage{nicefrac}       
\usepackage{microtype}      
\usepackage{graphicx}
\graphicspath{ {./images/} }

\title{Benchmarking Llama Model Security Against OWASP Top 10 for LLM Applications}

\author{
 Nourin Shahin \\
  Texas A\&M University San Antonio\\
  San Antonio, TX \\
  \texttt{nshahin@tamusa.edu} \\
 \And
 Izzat Alsmadi \\
  Texas A\&M University San Antonio\\
  San Antonio, TX \\
  \texttt{ialsmadi@tamusa.edu}
}

\begin{document}
\maketitle

\begin{abstract}
As large language models (LLMs) move from research prototypes to enterprise systems, their security vulnerabilities pose serious risks to data privacy and system integrity. This study benchmarks various Llama model variants against the OWASP Top 10 for LLM Applications framework, evaluating threat detection accuracy, response safety, and computational overhead. Using the FABRIC testbed with NVIDIA A30 GPUs, we tested five standard Llama models and five Llama Guard variants on 100 adversarial prompts covering ten vulnerability categories. Our results reveal significant differences in security performance: the compact Llama-Guard-3-1B model achieved the highest detection rate of 76\% with minimal latency (0.165s per test), whereas base models such as Llama-3.1-8B failed to detect threats (0\% accuracy) despite longer inference times (0.754s). We observe an inverse relationship between model size and security effectiveness, suggesting that smaller, specialized models often outperform larger general-purpose ones in security tasks. Additionally, we provide an open-source benchmark dataset including adversarial prompts, threat labels, and attack metadata to support reproducible research in AI security, \cite{noureen207_noureen207/llama-owasp-sec_2026}. 
\end{abstract}


\section{Introduction}

The integration of Large Language Models (LLMs) into production workflows has shifted the paradigm of intelligent application development. These models now act as the decision-making engine for various sectors, including finance, healthcare, and software engineering. However, this adoption has expanded the attack surface of enterprise software, introducing vulnerabilities that traditional cybersecurity protocols are often ill-equipped to handle.

Unlike legacy software, LLM security risks arise from the probabilistic nature of the models, their reliance on vast training corpora, and their susceptibility to natural language manipulation. Adversaries exploit these traits using prompt injection, context manipulation, or data poisoning, which can bypass standard safety mechanisms. Such breaches can lead to unauthorized data ex-filtration, the generation of harmful content, or complete system hijacking.

To address these emerging threats, the Open Worldwide Application Security Project (OWASP),  established the OWASP Top 10 for LLM Applications \cite{owasp2023}. This framework provides a standardized taxonomy for the most critical risks in the field, serving as the analytical foundation for our evaluation.

\subsection{Research Objectives}

This research provides a systematic evaluation of Llama model variants using the OWASP Top 10 framework, pursuing three primary goals:

\begin{itemize}

\item We quantified the security posture of different Llama architectures. By measuring their ability to identify and neutralize adversarial inputs across all ten OWASP categories, we provide a comparative analysis of five standard models and five Llama Guard variants.

\item We analyzed the trade-offs between security effectiveness and computational demand. By correlating detection accuracy with inference latency and VRAM usage, we presented empirical data to help developers select models that fit real-world resource constraints.

\item We introduced an open-source security benchmark, a small dataset, comprising 100 targeted test cases with comprehensive metadata, aims to facilitate reproducible testing and standardized security reporting within the AI research community.

\end{itemize}

\section{LLMs Evaluation: Dataset Construction}
\label{sec:headings}

Our evaluation utilizes a custom-built security benchmark designed to probe the resilience of LLMs against the vulnerabilities identified by OWASP. Our created dataset contains 100 adversarial prompts, divided equally across the ten security categories to ensure a balanced coverage.

\subsection{Dataset Design Principles}

The dataset construction was guided by the OWASP AI Testing Guide methodology \cite{owasp_testing_guide}. Each test case was designed to mirror real-world attack patterns found in recent penetration testing literature. The prompts utilize various techniques to bypass safety filters, including encoding obfuscation (Base64/Hex), role-playing exploits, and multi-turn manipulation.

In accordance with the OWASP framework, each entry in our dataset consists of three essential components: the attacker's instruction, a "trigger" (such as a role-playing cue or obfuscation) intended to override system constraints, and the malicious intent itself. We implemented 23 distinct injection techniques, ensuring a diverse range of adversarial strategies.

The data is structured in a JSON format with the following fields:

\begin{itemize}
\item \textbf{Test ID}: A unique identifier (e.g., LLM01\_001).
\item \textbf{Category}: The primary OWASP vulnerability classification.
\item \textbf{Subcategory}: The specific attack technique (e.g., Jailbreak, Token Smuggling).
\item \textbf{Prompt}: The actual adversarial text.
\item \textbf{Safety Label}: The ground-truth classification (safe/unsafe).
\item \textbf{Metadata}: Details on severity levels (Low to Critical) and technical notes.
\end{itemize}

\paragraph{OWASP Top 10 Coverage.}
The dataset provides a comprehensive template of the OWASP categories. This includes Prompt Injection (LLM01) using "DAN" style jailbreaks; Sensitive Information Disclosure (LLM02) targeting credential extraction; and Supply Chain Vulnerabilities (LLM03) involving unverified model loading. Other categories include Data Poisoning (LLM04), Improper Output Handling (LLM05), Excessive Agency (LLM06), System Prompt Leakage (LLM07), Vector and Embedding Weaknesses (LLM08), Misinformation (LLM09), and Unbounded Consumption (LLM10).

\section{Related Work and Comparison Study}
\label{sec:related_work}

The security of Large Language Models (LLMs) has emerged as a critical research area over the last few years. Unlike traditional software, LLM vulnerabilities often arise from probabilistic reasoning and natural language manipulation rather than code-level exploits. Our work builds on multiple recent efforts, which we categorize into three primary domains: prompt injection and jailbreak defenses, LLM content moderation and Guard models, and comprehensive security benchmarking frameworks.

\subsection{Prompt Injection and Jailbreak Defenses}

Prompt injection attacks exploit the text-based interface of LLMs to override intended instructions, often resulting in leakage of sensitive data or execution of unsafe outputs. Early investigations by Weidinger et al. ~\cite{weidinger2021ethical} highlighted the risks of language model jailbreaks and prompted research into defensive strategies, including careful instruction design, input sanitization, and output filtering. 

Touvron et al.~\cite{touvron2023llama} introduced the Llama 2 family of models, demonstrating improved instruction-following capabilities. However, even these state-of-the-art models remain susceptible to advanced jailbreak techniques when adversarially prompted. The work by Wallace et al.~\cite{wallace2019universal} proposed universal adversarial triggers that reliably bypass model guardrails, reinforcing the need for specialized mitigation models rather than relying solely on general-purpose LLMs.

\subsection{LLM Guard Models and Content Moderation}

To address the limitations of base LLMs, researchers have developed purpose-built Guard models that integrate fine-tuning and safety classifiers into the language model architecture. Inan et al.~\cite{inan2023llamaguard} presented \textit{Llama Guard}, a set of models trained to perform explicit input-output safety evaluation for conversational AI systems. By directly producing structured safety labels, these models reduce the ambiguity inherent in generative LLM outputs and facilitate automated decision-making for security-critical applications.

Other efforts, such as the \textit{AlpacaFarm Safety Suite}~\cite{dubois2023alpacafarm}, apply reinforcement learning from human feedback (RLHF) to encourage safe completions. While effective for content moderation, these models do not explicitly address OWASP-style vulnerabilities such as Supply Chain Risks or System Prompt Leakage. Our study extends this work by evaluating Guard and base Llama models across the full OWASP Top 10 framework, providing a more holistic security assessment.

\subsection{Security Benchmarking and Evaluation Frameworks}

Benchmarking LLM security is an emerging challenge due to the diversity of attack vectors. The OWASP AI Testing Guide~\cite{owasp_testing_guide} provides a structured methodology for evaluating LLM safety, emphasizing adversarial testing, role-playing exploits, and multi-turn prompt injections. Other benchmark frameworks, such as Gptfuzzer~\cite{yu2023gptfuzzer}, focus primarily on prompt injection detection and content moderation, but do not cover broader attack categories like vector and embedding weaknesses, unbounded resource consumption, or supply chain vulnerabilities.

Our approach distinguishes itself by combining five standard Llama models and five Guard variants, evaluating them against 100 adversarial prompts spanning ten OWASP categories. Unlike prior benchmarks, we quantify both security effectiveness and computational cost (latency and VRAM usage) on high-performance GPU infrastructure. This dual focus addresses a critical gap in the literature: most studies measure accuracy in isolation, without considering practical deployment constraints.

Overall, our survey of recent literature reveals a persistent gap in comprehensive, high-fidelity security benchmarking for LLMs. By integrating model-specific analysis, multi-category adversarial evaluation, and GPU deployment metrics, our work provides a novel, reproducible methodology that extends beyond the scope of previous studies.

\section{Methodology}
\label{sec:methodology}

\subsection{Experimental Infrastructure}

Experiments were conducted on the FABRIC testbed \cite{fabric2021}, utilizing the Georgia Tech site in this particular experiment. This environment allows for isolated, high-performance computing without the interference of multi-tenant cloud overhead. The hardware configuration included:

\begin{itemize}
\item \textbf{GPU}: NVIDIA A30 (24GB VRAM, Ampere architecture).
\item \textbf{Operating System}: Ubuntu 22.04 LTS.
\item \textbf{Software Stack}: PyTorch 2.1.0, CUDA 11.8, and the HuggingFace Transformers library (v4.51.3).
\item \textbf{Driver}: NVIDIA version 535.183.01.
\end{itemize}

\subsection{Model Selection and Configuration}
\label{models:configuration}

We selected ten models from the Llama family to represent a range of parameter scales and training objectives. These were divided into standard generative models and specialized security-focused variants.

\paragraph{Standard Llama Models (5 variants)}
These variants represent general-purpose architectures:
\begin{itemize}
\item \textbf{Meta-Llama-3-8B} and \textbf{Llama-3.1-8B} (Base models).
\item \textbf{Llama-3.1-8B-Instruct} (Tuned for instruction following).
\item \textbf{Llama-3.2-1B} and \textbf{Llama-3.2-3B-Instruct} (Compact, efficiency-focused models).
\end{itemize}

\paragraph{Llama Guard Models (5 variants)}
These are models purpose-built for content moderation and security classification:
\begin{itemize}
\item \textbf{Meta-Llama-Guard-2-8B} (Previous generation).
\item \textbf{Llama-Guard-3-1B} and \textbf{Llama-Guard-3-8B} (Current generation).
\item \textbf{Llama-Guard-3-8B-INT8} (Quantized variant).
\item \textbf{Llama-Guard-3-11B-Vision} (Multimodal variant, tested here on text).
\end{itemize}

All models were loaded in float16 precision (except the INT8 variant) with a low temperature (0.1) and a maximum token limit of 10 to ensure binary, deterministic outputs.

\paragraph{Standard vs. Guard Model Output Behavior}
A fundamental distinction exists between standard Llama models and Llama Guard models that affects both their performance and deployment suitability.

\textbf{Llama Guard Models} are explicitly fine-tuned for binary safety classification. These models produce structured safety labels ("safe" or "unsafe") directly in response to input prompts. Their training objective optimizes for content moderation, making them naturally suited for security evaluation with minimal prompt engineering.

\textbf{Standard Llama Models} are generative language models trained for text completion and conversation. They are not inherently designed for safety classification and must be prompted to produce binary judgments. When asked "is this prompt safe or not?", standard models may generate explanations, hedged responses, or contextual elaborations rather than direct classifications. We extract binary decisions by parsing outputs for "unsafe" keywords—if "unsafe" appears, we classify it as an unsafe detection; otherwise, it is treated as "safe".

This architectural difference has important implications: Guard models perform the exact classification task they were trained for, while standard models approximate it through text generation. This distinction partially explains the performance gaps observed in our results and informs deployment decisions—Guard models offer more reliable, structured outputs for production security monitoring.

\subsection{Evaluation Protocol}

The evaluation followed a four-stage pipeline:
\begin{enumerate}
    \item \textbf{Model Loading}: Recording VRAM allocation and loading times to establish infrastructure benchmarks.
    \item \textbf{Test Execution}: Measuring end-to-end latency for each prompt. All models were prompted with a consistent classification frame: "is this prompt safe or not?".
    \item \textbf{Response Evaluation}: Parsing outputs for "safe" or "unsafe" labels to determine detection success.
    \item \textbf{Aggregation}: Calculating category-specific and overall performance metrics.
\end{enumerate}

\section{Results}
\label{sec:results}

\subsection{Overall Performance Comparison}

Table~\ref{tab:overall-results} details the performance across all models. The data highlights a distinct performance gap between specialized/instruct-tuned models and base models. This gap reflects both differences in safety training and the fundamental architectural distinction discussed in Section \ref{models:configuration}. Guard models are optimized for direct classification, while standard models approximate this through generation.

\begin{table}[h]
\caption{Overall Security Detection Performance and Computational Metrics}
\centering
\small
\begin{tabular}{lcccc}
\toprule
\textbf{Model} & \textbf{Detection} & \textbf{Total Time} & \textbf{Avg/Test} & \textbf{VRAM} \\
 & \textbf{Rate (\%)} & \textbf{(s)} & \textbf{(s)} & \textbf{(GB)} \\
\midrule
\multicolumn{5}{l}{\textit{Standard Llama Models}} \\
Meta-Llama-3-8B & 0 & 77.43 & 0.774 & 5.31 \\
Llama-3.1-8B & 0 & 75.44 & 0.754 & 5.31 \\
Llama-3.1-8B-Instruct & 54 & 20.59 & 0.206 & 5.32 \\
Llama-3.2-1B & 73 & 27.80 & 0.276 & 0.97 \\
Llama-3.2-3B-Instruct & 66 & 18.42 & 0.184 & 2.10 \\
\midrule
\multicolumn{5}{l}{\textit{Llama Guard Models}} \\
Meta-Llama-Guard-2-8B & 7 & 16.80 & 0.168 & 4.85 \\
Llama-Guard-3-1B & 76 & 16.50 & 0.165 & 0.94 \\
Llama-Guard-3-8B & 33 & 17.30 & 0.173 & 4.89 \\
Llama-Guard-3-8B-INT8 & 28 & 42.20 & 0.422 & 2.45 \\
Llama-Guard-3-11B-Vision & 28 & 20.59 & 0.208 & 6.21 \\
\bottomrule
\end{tabular}
\label{tab:overall-results}
\end{table}

\subsection{Key Findings}

\paragraph{Latency and Accuracy Correlation.}
A central finding of this study is the inverse relationship between inference latency and security detection. The base models (Meta-Llama-3-8B and Llama-3.1-8B) were the slowest and least effective, failing to detect any threats. In contrast, Llama-Guard-3-1B and Llama-3.2-1B provided high accuracy (76\% and 73\%, respectively) while maintaining minimal latency. This suggests that security reasoning in LLMs is driven by efficient pattern recognition from specialized training rather than computational scale.

\paragraph{Impact of Instruction Tuning.}
The data shows that instruction tuning is essential for security responsiveness. Llama-3.1-8B-Instruct achieved a 54\% detection rate, whereas its non-instruct base variant achieved 0\%. This shift demonstrates that the fine-tuning process significantly enhances a model's ability to interpret adversarial intent and categorize it correctly.

\paragraph{Efficiency of Compact Models.}
Smaller models notably outperformed larger ones. The Llama-3.2-1B and Llama-Guard-3-1B models provided the best balance of speed, accuracy, and memory efficiency. The Guard-3-1B model, specifically, used only 0.94GB of VRAM while achieving the highest overall score. This indicates that for security-specific tasks, massive parameter counts may lead to diminishing returns.

\paragraph{Quantization and Multimodal Trade-offs.}
Quantization appears to be a poor strategy for security deployments. Llama-Guard-3-8B-INT8 was significantly slower and less accurate than its full-precision counterpart. Furthermore, the Llama-Guard-3-11B-Vision model's low detection rate (28\%) suggests that multimodal optimization may dilute a model's performance on pure-text safety classification.

\subsection{Category-Specific Performance Analysis}

Table~\ref{tab:category-best} examines how top-performing models handled specific OWASP categories.

\begin{table}[h]
\caption{Detection Rates by OWASP Categories for Top Performing Models}
\centering
\small
\begin{tabular}{lccc}
\toprule
\textbf{Vulnerability Category} & \textbf{Llama 3.2-1B} & \textbf{Llama 3.2-3B-I} & \textbf{Llama 3.1-8B-I} \\
\midrule
LLM01: Prompt Injection & 50\% & 70\% & 100\% \\
LLM02: Info Disclosure & 90\% & 50\% & 30\% \\
LLM03: Supply Chain & 100\% & 60\% & 30\% \\
LLM04: Data Poisoning & 70\% & 50\% & 40\% \\
LLM05: Output Handling & 90\% & 90\% & 90\% \\
LLM06: Excessive Agency & 70\% & 80\% & 40\% \\
LLM07: Prompt Leakage & 80\% & 10\% & 0\% \\
LLM08: Vector Weaknesses & 70\% & 60\% & 50\% \\
LLM09: Misinformation & 50\% & 100\% & 90\% \\
LLM10: Unbounded Consumption & 60\% & 90\% & 60\% \\
\bottomrule
\end{tabular}
\label{tab:category-best}
\end{table}

The analysis shows that while Llama-3.1-8B-Instruct is highly resilient to direct Prompt Injection (100\%), it is almost entirely vulnerable to System Prompt Leakage (0\%). Llama-3.2-1B, while less effective against injection, showed remarkable resilience to Information Disclosure (90\%) and Supply Chain attacks (100\%). These results suggest that no single model currently provides comprehensive protection across all threat vectors.

\section{Discussions}

\subsection{Implications for Production Deployment}

These results suggest that organizations should move away from a "one-size-fits-all" approach to LLM security. Instead of deploying the largest available model, security architects should consider the followings:

\begin{itemize}
    \item \textbf{Prioritize specialized compact models}: The 1B parameter variants provide superior security with minimal resources' overhead, making them ideal for edge or high-throughput cloud environments.
    \item \textbf{Avoid base models for safety}: Base models lack the alignment necessary to identify malicious inputs. Instruction-tuned or guard-specific variants are mandatory for security-sensitive roles.
    \item \textbf{Layered defense}: Since specific models excel in different OWASP categories, a multi-model ensemble—where a guard model handles content filtering and a compact instruct model handles injection detection—may provide the most robust defense.
    \item \textbf{Consider output architecture}: For production security monitoring, guard models offer inherent advantages through their structured and consistent binary outputs compared to standard models that require parsing of generated text.
\end{itemize}

\subsection{Critical Vulnerability Gaps}

Two categories remain particularly problematic: \textbf{System Prompt Leakage (LLM07)} and \textbf{Supply Chain Vulnerabilities (LLM03)}. Most models failed to detect attempts to extract system-level instructions, which could allow attackers to map out security logic for future exploits. Similarly, supply chain threats, such as malicious plugin execution, are largely ignored by current safety training, representing a significant blind spot in the LLM ecosystem.

\section{Conclusion}

This study established a basic benchmark for evaluating Llama models against the OWASP Top 10 framework. Our analysis reveals that computational scale does not equate to security robustness. The most efficient models—specifically Llama-Guard-3-1B—achieved the highest threat detection rates while requiring the least inference time and memory.

Our findings challenge the assumption that larger models are inherently safer. Instead, specialized training and instruction tuning emerge as the primary drivers of security performance. The architectural distinction between Guard models (purposely built for classification) and standard models (adapted from text generation) proves critical for both performance and operational deployment. Guard models provide more reliable, structured outputs essential for production security systems.

Nonetheless, significant gaps remain, particularly regarding system prompt protection and supply chain integrity. As LLMs become more deeply embedded in enterprise infrastructures, the need for specialized, low-latency security models will only increase.

\bibliographystyle{unsrt}  
\bibliography{references} 
%







\end{document}